\documentclass[amsmath,a4paper,11pt]{article}
\pdfoutput=1 

\usepackage{jheppub} 

\usepackage[T1]{fontenc} 
\usepackage{tcolorbox}

\usepackage{slashed}

\title{\bf  Defect QED:  Dielectric without a Dielectric, Monopole without a Monopole}

\author[a]{Gianluca Grignani,}
\author[b]{Gordon W. Semenoff}

\affiliation[a]{Dipartimento di Fisica e Geologia, Universit\`a  di Perugia, I.N.F.N. Sezione di Perugia,
Via Pascoli, I-06123 Perugia, Italy}
\affiliation[b]{Department of Physics and Astronomy, 
University of British Columbia, 
6224 Agricultural Road, 
Vancouver, British Columbia, Canada V6T 1Z1}
 
\emailAdd{gianluca.grignani@unipg.it}
\emailAdd{gordonws@phas.ubc.ca}

\abstract{We study a class of defect quantum field theories where the quantum field theory in the 3+1-dimensional bulk is a free photon and charged matter and the interactions of the photons with the charges occur entirely on a 2+1-dimensional defect.  We observe that at the fully quantum level, the effective action of such a theory is still a defect field theory with free photons propagating in the bulk and the nonlinearities in the quantum corrections to the Maxwell equations confined to the defect.  We use this observation to show that the defect field theory has interesting electromagnetic properties. The electromagnetic fields sourced by static test charges are attenuated as if the bulk surrounding them were filled with a dielectric material. This is particularly interesting when the observer and test charge are on opposite sides of the defect.  Then the effect is isotropic and it is operative even in the region near the defect.  If the defect is in a time reversal violating state, image charges have the appearance of electrically and magnetically charged dyons. We present the example of a single layer in a quantum Hall state.  We observe that the charge screening effect in charge neutral graphene should be significant, and even more dramatic  when the layer is in a metallic state with mobile electrons.    }

\begin{document} 

\maketitle
\flushbottom

 \section{Introduction}
 
 It has recently become apparent that much can be learned  about the structure of quantum field theories  by studying their behaviour in the presence
 of boundaries and defects.  Moreover, as we shall observe, the behaviour itself could embody  interesting effects which could be important in physical applications of the quantum field theories.  
 There has recently been an effort to understand a boundary
  electrodynamics model where a photon field resides in the bulk of 3+1-dimensional space-time and charged matter occupies a 2+1-dimensional boundary and interacts with the photon through interactions which are localized on the boundary \cite{Herzog:2018lqz}-\cite{DiPietro:2019hqe}.  This scenario has been used as an example of a conformal field theory with an exactly marginal deformation, the electric charge, which is conjectured to have vanishing beta function.  A perturbative computation suggests that, in the weak coupling regime,  the boundary central charges ``$b_1$'' and ``$b_2$'' and other data of the conformal field theory  vary continuously with the strength of this coupling.   
   
  In this paper, we wish to point out that a defect conformal field theory which is free field theory in the bulk has some  special properties which occur simply due to its geometry.  We will use a defect rather than boundary quantum field theory.  Like the boundary field theory, the defect theory has a free photon in the bulk which is coupled to the electromagnetic currents of  interacting charged fields residing on the defect.   We will consider an infinite, flat, 2+1-dimensional defect which bisects 3+1-dimensional Minkowski spacetime (which we shall call the ``bulk'') into two disjoint regions.  
  In this context, the main difference between the boundary and the defect conformal field theory is in the boundary conditions.  For a boundary field theory, the electromagnetic field strength obeys the boundary condition
\begin{align}\label{boundary_bc}
F^{\mu \perp} (x\to{\rm boundary})=J^\mu_{\rm boundary}(x)
\end{align}
whereas for a defect field theory it obeys the boundary condition
\begin{align}\label{defect_bc}
F^{\mu \perp} (x\to{\rm defect}^+)-F^{\mu \perp} (x\to{\rm defect}^-)=J^\mu_{\rm defect}(x)
\end{align}
A comprehensive discussion of bulk Abelian gauge theories with the boundary condition (\ref{boundary_bc}) 
and other boundary conditions related by SL(2,Z) duality \cite{Witten} \cite{Gaiotto} is given in reference \cite{DiPietro:2019hqe}.
Use of the defect boundary condition on the other hand has a long history, for example see the discussion in reference \cite{Gorbar:2001qt}.
In this paper, we shall use the defect boundary condition.  Much of our discussion could easily be adapted to the boundary field theory case.  When we consider the case of a conformal field theory,  the conformal symmetry is the $SO(3,2)$ conformal group of the defect.  

Although most of our results depend only on the defect current-current correlation function and have a wider degree of generality, we will consider the concrete example of the 
defect quantum field theory whose (Euclidean signature time) correlation functions are computed by inserting operators into the  functional integral
\begin{align}
 Z[j]=\frac{
\int [d\psi d\bar\psi dA_\mu]e^{-S[A,\psi,\bar\psi]+\int d^4x j^\mu(x) A_\mu(x)} 
 }{\int [d\psi d\bar\psi dA_\mu] e^{-S[A,\psi,\bar\psi]}
}
\label{partition_function}
\end{align}
where the action has $4$-dimensional bulk and $3$-dimensional defect components
   \begin{align}
\label{action}   &S[A,\psi,\bar\psi]=S_4+S_3=\int d^4x~\left\{  {\mathcal L}_4(x) + \delta(z) {\mathcal L}_3(x) \right\}
 \\
 \label{lagrangian4}
  &{\mathcal L}_4(x)  = \frac{1}{4}F_{\mu\nu}(x) F^{\mu\nu}(x) +i \frac{\theta}{4\pi}\epsilon^{\mu\nu\rho\sigma}F_{\mu\nu}(x)F_{\rho\sigma}(x) 
\\  \label{lagrangian3}
 & {\mathcal L}_3(x)=
 -i\bar\psi(x)\gamma^aD_a \psi(x)  +i \frac{\tilde\kappa}{4\pi} \epsilon_{abc}A_a(x)\partial_b A_c(x)
\end{align}
Here,  $\hbar=1=c$,  $\mu,\nu,...$ index four-vectors, for example  $x^\mu=(t,x,y,z)$.  The defect is located at $z=0$. The indices $a,b,...$ denote three-vectors which lie along the defect world-volume, for example, $x^a=(t,x,y)$.  Although up and down indices are not needed in Euclidean space,  in order to make it easier for the reader to convert our formulae to Minkowski space, we will use up and down indices and we will adhere to the usual summation conventions. 

The U(1)  vector gauge field is $A_\mu(x)$, the defect covariant derivative is $D_a\equiv\partial_a-ieA_a$ and the field strength is $F_{\mu\nu}=\partial_\mu A_\nu-\partial_\nu A_\mu$. The field
$\psi(x)$ is a two-component spinor of the Euclidean $SO(3)$ Lorentz group of the defect and we shall consider $N$ species of these spinors so that the
field theory has a global $U(N)$ symmetry. The Dirac matrices have the algebra $\left\{\gamma^a,\gamma^b\right\}=2\delta^{ab}$.  The three and four dimensional Lagrangian densities written above contain all of the marginal local operators that can be formed with the vector and spinor fields which are  gauge invariant and Lorentz invariant. 

The defect Chern-Simons term could be replaced by a discontinuity of the theta-angle in the topological term $i \frac{\theta}{4\pi}\epsilon^{\mu\nu\rho\sigma}F_{\mu\nu}(x)F_{\rho\sigma}(x) $ which would then have $\theta$ jumping by $\tilde\kappa$ as one crosses the defect.  The Chern-Simons term violates parity and time reversal invariance.  If these symmetries are imposed, both $\tilde\kappa$ and $\theta$ should be set to zero.  Since $\theta$ will not play a role in the following, we will hereafter set it to zero, and retain the Chern-Simons term when we discuss materials which violate time reversal symmetry.\footnote{We will not restrict our consideration to compact QED, where the theta term plays a more important role.  For most of our results, there is very little difference between the compact and non-compact theories.}

There are a few interesting relevant deformations of this theory.   One would be a Fermion mass term $im\bar\psi T\psi$ where $T$ is a generator of the U(N) symmetry (possibly the unit matrix). A mass term must violate either parity and time reversal invariance or the SU(N)$\subset$U(N) global symmetry.   In order to preserve parity and time reversal invariance, there would have to be another generator, say $S$,  of  U(N) with the property $S^\dagger T+TS=0$, so that the parity and time-reversal transformations would be augmented by $\psi\to S\psi$.  On the other hand, a U(N) invariant $im\bar\psi\psi$ mass term violates parity and time reversal.   The Fermion mass deformations gap the spectrum of the Fermions on the defect and they render the defect an insulator.    Moreover, when the Chern-Simons term is present, there is no symmetry which protects the Fermions from obtaining a U(N) symmetric mass term and it must be tuned to zero.  This can be done consistently order by order in perturbation theory \cite{Chen:1992ee}-\cite{Chen:1990sc}.  Another important relevant deformation is a chemical potential term $$\delta   {\mathcal L}_3(x) = 
i\mu \bar\psi(x)\gamma^0\psi(x)$$ which can be used to control the Fermion density.  We will use it to make a metallic state of the defect.   

 The interaction $e\bar\psi (x)\gamma^a A_a \psi(x)$ is thought to be  exactly marginal, at least for those values of the coupling constant $e$ which are in the perturbative regime.  As well, the $\theta$-term and the Chern-Simons terms have vanishing perturbative beta-functions and this field theory is thought to be a defect conformal field theory for any values of these parameters which are in the neighborhood of weak coupling.  We should point out that strong interactions in this theory are thought to result in the generation of a mass gap for the Fermions \cite{Pisarski:1984dj} \cite{Appelquist:1986fd}
 driven either by strong Coulomb interactions or by four-Fermi interactions which can become relevant at strong enough coupling.  In our discussions of defect conformal field theory, we will assume that the couplings are weak enough that none of these occur.

 In the context of the defect field theory that we have outlined, 
we shall be interested in the effective field theory of the photon.  
The essential observation that we make here  will be that the fully quantum corrected effective field theory  of the photon is still a defect field theory with a free photon in the $3+1$-dimensional bulk and interactions confined to the defect.  This observation depends only on the fact that, at the tree level, the bulk is described by free field theory and that the dynamical currents are confined to the defect, so this should hold in other interesting field theories as well and, in particular, it does not depend on conformal or Lorentz invariance.    With this observation, we will examine the weak field response of the defect field theory to the presence of an external electric charge and current density, which can be positioned either in the bulk or on the defect.   We will find that the presence of the defect alters the electromagnetic field due to the charge and current density in an interesting way.  In the presence of the charge or current density, the presence of the defect induces an image charge or current.     The position of the image is always on the side of the defect which is opposite to the observer, irrespective of the position of the charge and current density.  If the charge and current density are on the defect, the position of the image coincides with their positions.  If the defect field theory violates time reversal, the image has magnetic charge as if it were a Dirac monopole.  The magnetic monopole charge generally does not obey the Dirac quantization condition.  We shall discuss how this quantization is avoided in this case.  Also, the charge is screened with the screening depending on the properties of the defect field theory.  If the defect is metallic, the charge is completely screened.  If it is semi-metallic, scale invariance as well as other symmetries lead to a screening which is isotropic and appears as if the spacetime in the bulk, outside of the defect, is filled with a dielectric substance.  Although we shall not discuss them,  our results could have interesting implications for graphene or other single layer materials.  

\section{The effective action of defect electrodynamics}

To begin, let us recall the definition of the effective field theory.
We begin with the classical field, the one-point function of the photon, $\alpha_\mu(x)$, which is induced by the presence of the classical current  $j_\mu(x)$ in the partition function in equation (\ref{partition_function}),  computed by
\begin{align}\label{a}
\alpha_\mu(x)~=~\frac{\partial}{\partial j^\mu(x)}\ln Z[j] 
\end{align}
In this equation, $x$ has support in the entire four-dimensional spacetime and it could be placed on the defect.   As usual, the index $\mu$ takes on
the values $(t,x,y,x)$. 
We then find the effective action for this classical field $\alpha_\mu(x)$ by taking a Legendre transform of the generating functional, 
\begin{align}\label{gamma}
&\Gamma[\alpha]~=~-~\ln Z[j]~+~\int d^4x~ j^\mu(x) \alpha_\mu(x)
\end{align}
 From the structure of the Legendre transformation, the classical field $\alpha_\mu(x)$ obeys an equation of motion 
\begin{align}\label{max}
\frac{\delta}{\delta \alpha_\mu(x)}~\Gamma[\alpha]= 
~j^\mu(x)
\end{align}
which are the quantum corrected (Euclidean signature) Maxwell equations.  They govern the response of the electromagnetic fields to the 
presence of an external charge and current density, $j^\mu(x)$.   Note that, unlike the electromagnetic currents of the defect fields, which necessarily reside on the defect, $j^\mu(x)$ can have support either on the defect or in the bulk. 

Our observation is that the classical field equation (\ref{max})  is still that of a defect  field theory in the sense that the non-linearities in that equation  reside entirely on the defect. 
  That is, the effective action $\Gamma[\alpha]$ has the form
\begin{align}\nonumber
\Gamma[\alpha]=&\int d^4x \frac{1}{4}\left[ \partial_\mu \alpha_\nu(x)-\partial_\nu \alpha_\mu(x)\right]\left[ \partial^\mu \alpha^\nu(x)-\partial^\nu \alpha^\mu(x)\right]+\\
 &+\sum_{n=2}^\infty\frac{1}{n!}\int d^3x_1\ldots d^3x_n\Gamma^{a_1\ldots a_n}(x_1,\ldots,x_n)\alpha_{a_1}(x_1)\ldots \alpha_{a_n}(x_n)
\label{gamma1} \end{align}
 The coefficients in the functional Taylor expansion are ($-1$ times) the one-photon-irreducible time-ordered correlation functions of the defect currents, $J^a (x) =- e\bar\psi(x)\gamma^a\psi(x)$,  
 \begin{align}\label{corr}
\Gamma^{a_1\ldots a_n}(x_1,\ldots,x_n)= - \left<0| {\mathcal T}J^{a_1}(x_1)\ldots J^{a_n}(x_n)|0\right>_{1PI}
\end{align}

To see why this is the case, we begin with 
the intermediate step of 
observing that the Dirac Fermion fields $\psi(x)$ and $\bar\psi(x)$ appear quadratically in the action (\ref{action}) and they can be integrated out of the functional integral (\ref{partition_function}) to get an intermediate effective action for the electromagnetic field, 
\begin{align} 
S_{\rm eff}[A]=&\int d^4xd^4y   \frac{1}{2}  A_a(x) \Delta^{-1}_{ab}(x,y) A_b(y)
-\ln\det\left[ -i\gamma^a \partial_a +e\gamma^aA_a(x) \right] \label{seff1}\\
=&\int d^4xd^4y   \frac{1}{2}  A_a(x) \Delta^{-1}_{ab}(x,y) A_b(y)+S_{\rm int}[A]
\label{seff1.5} \\
&S_{\rm int}[A]
=\sum_{n=2}^\infty\frac{1}{n!}\int d^3x_1\ldots d^3x_n\tilde\Gamma^{a_1\ldots a_n}(x_1,\ldots,x_n)A_{a_1}(x_1)\ldots A_{a_n}(x_n)
\label{seff2}\\ \label{delta}
&\Delta^{-1}_{ab}(x,x')=\left[-(\partial_z^2+\partial^c\partial_c)\delta_{ab}+\partial_a\partial_b\right]\delta^4(x-x') 
 \end{align} 
 where the expression in equation (\ref{seff2}) is the functional Taylor expansion of the Fermion determinant which is the last term 
 in equation (\ref{seff1}) and which we call the interaction action, $S_{\rm int}[A]$, the last term in equation (\ref{seff1.5}).  
 The coefficients in the expansion in equation (\ref{seff2}), 
$\tilde \Gamma^{a_1\ldots a_n}(x_1,\ldots,x_n)$, are  ($-1$ times) the one-photon irreducible   $n$-point defect current correlation functions, that is, the functions which appear in equation  (\ref{corr}), but computed in the one-loop approximation.  In the following, we will not worry about renormalization of the field theory.  We assume that all of our statements hold for a cutoff, ultraviolet regulated theory and that counter-terms can be chosen so as to remove the singularities from the  effective action, $\Gamma[\alpha]$, once it is found.  Renormalization of this model in the leading orders of perturbation theory, as well as some interesting one and higher loop computations have been performed in a series of interesting papers \cite{teber_1}-\cite{teber_4}.  We have also fixed the $A_z(x)=0$ gauge, which affects only the bulk terms
where it results a Gaussian term for the remaining vector fields, $A_a(x)$ with the invertible quadratic form $\Delta^{-1}_{ab}(x,x')$. The effective action is now given by the functional integral expression
\begin{align}\label{fint1}
\Gamma[\alpha]= - \ln\left\{ \int [d  A_a]  e^{-S_{\rm eff}[ A_a ] + \int (A_a-\alpha_a) j^a} \right\}+ {\rm constant}
\end{align} 

Then, the remaining task is to integrate over the vector field $A_a(x)$. We are not able to do this.  However, we can study some of
the properties of the integral.  To facilitate this, we  do the change of variables in the
functional integral $A_a(x)=\alpha_a(x)+\tilde A_a(x)$ where $\alpha_a(x)$ is the one-point function of the field $A_a(x)$ that is defined in equation (\ref{a}) and $\tilde A_a(x)$ is the new functional integration variable.   We are left with the functional integral expression for the effective action
\begin{align}\label{fint}
\Gamma[\alpha]= - \ln\left\{ \int [d\tilde A_a]  e^{-S_{\rm eff}[\alpha_a + \tilde A_a ] + \int \tilde A_a j^a} \right\}+ {\rm constant}
\end{align} 
 Since the one-point function for $A_a(x)$ is equal to $\alpha_a(x)$, 
the one-point function of $\tilde A_a(x)$ must vanish.   This functional integral  (\ref{fint}) determines $\Gamma[\alpha]$ when $j^a(x)$ is determined to be that functional of $\alpha_a(x)$ so that the
one-point function of $\tilde A_a(x)$ indeed vanishes.  This is equivalent to the equation of motion for $\alpha_a(x)$ in equation (\ref{max}).  To see this directly, consider
a combination of equations (\ref{max}) and (\ref{fint}),
\begin{align*}
&0=- \frac{\delta\Gamma[\alpha] }{\delta \alpha_a(x)}+ j^a(x)
= \frac{\delta }{\delta \alpha_a(x)}\ln \left\{ \int [d\tilde A] 
  e^{-S_{\rm eff}[\alpha + \tilde A ]+\int \tilde A j} \right\}+j^a(x) \\
&=\frac{  \int [d\tilde A] 
\left[  
\frac{\delta}{\delta \alpha_a(x)} + j^a(x)\right]  e^{-S_{\rm eff}[\alpha + \tilde A ]+\int \tilde A j}  }{  \int [d\tilde A]  e^{-S_{\rm eff}[\alpha + \tilde A ]+\int \tilde A j}  } \\
&=\frac{  \int [d\tilde A]
\left[   
\frac{\delta}{\delta \tilde A_a(x)} +\int d^4y \frac{\delta j^b(y)}{\delta \alpha_a(x)}\tilde A_b(y)\right]  e^{ - S_{\rm eff}[\alpha + \tilde A ]+\int \tilde A j}  }{  \int [d\tilde A]  e^{ - S_{\rm eff}[\alpha + \tilde A ]+\int \tilde A j}  }
\nonumber \\&=  \int d^4y  \frac{\delta j^b(y)}{\delta \alpha_a(x)}~\frac{   \int [d\tilde A]
 e^{ - S_{\rm eff}[\alpha + \tilde A ]+\int \tilde A j}   \tilde A_b(y) }{  \int [d\tilde A]  e^{ - S_{\rm eff}[\alpha + \tilde A ]+\int \tilde A j}  }
 \end{align*}
where we have assumed that the functional integral of the functional derivative of $  e^{-S_{\rm eff}[\alpha + \tilde A ]+\int \tilde A j}$ vanishes.  We
assume that the kernel $ \frac{\delta j^b(y)}{\delta \alpha_a(x)}$, which is proportional to the inverse of the two-point function, is invertible.  Then, from the above, we conclude that, consistent with our construction, 
 \begin{align*}
\frac{  \int [d\tilde A]
 e^{ - S_{\rm eff}[\alpha + \tilde A ]+\int \tilde A j}   \tilde A_b(y) }{  \int [d\tilde A]  e^{ - S_{\rm eff}[\alpha + \tilde A ]+\int \tilde A j}  }=0
 ~~\iff~~~\frac{\delta}{\delta \alpha_a(x)}~\Gamma[\alpha] = j^a(x)
\end{align*}

We can also easily show that $\Gamma[\alpha]$ is   irreducible.  The definition of an irreducible functional is that it cannot be rendered a sum of two functionals of $\alpha$ by removing a single internal two-point-function.  (This could either be the bare two-point function $\Delta_{ab}(x,y)$ or the full
two-point function with interactions included.   For simplicity, we will use $\Delta_{ab}(x,y)$.) To proceed, let us consider the functional integral (\ref{fint}) 
but with $\Delta^{-1}_{ab}(x,y)$ replaced, for the moment, with  an arbitrary invertible kernel, rather than the one that is specified in equation (\ref{delta}).
Then, we can obtain a sum of terms, each one of which is the expression that is obtained by removing a different one of the 2-point functions $\Delta_{ab}(x,y)$ from  $\Gamma[\alpha]$ 
by 
taking the functional derivative  (while holding $\alpha_a(x)$ and $j_a(x)$ fixed)

\begin{align}
 \left. 2\frac{\delta}{\delta \Delta^{ab}(x,y)}\Gamma[\alpha]\right|_{j,\alpha}  
=\int d^4dd^4z'
 \Delta^{-1}_{bc}(y,z)\frac{ \int [d\tilde A] e^{-S_{\rm eff}[\tilde A,\alpha]}~2\frac{\delta S_{\rm eff}[\alpha+\tilde A] }{\delta \Delta^{-1}_{cd}(z,z')}
  }  { \int [d\tilde A] e^{ - S_{\rm eff}[\tilde A,\alpha]}} 
\Delta^{-1}_{da}(z',x)
\\
=\int d^4dd^4z'
 \Delta^{-1}_{bc}(y,z)\frac{ \int [d\tilde A] e^{ - S_{\rm eff}[\tilde A,\alpha]}\left(\alpha_c(z)+\tilde A_c(z)\right)\left(\alpha_d(z')+\tilde A_d(z')\right) }  { \int [d\tilde A] e^{ - S_{\rm eff}[\tilde A,\alpha]}} 
\Delta^{-1}_{da}(z',x) \\
=\int d^4dd^4z'
 \Delta^{-1}_{bc}(y,z) \left[ \alpha_c(z) \alpha_d(z')+
\frac{ \int [d\tilde A] e^{ - S_{\rm eff}[\tilde A,\alpha]}\tilde A_c(z)\tilde A_d(z') }  { \int [d\tilde A] e^{ - S_{\rm eff}[\tilde A,\alpha]}} \right]
\Delta^{-1}_{da}(z',x)
\label{prf}
\end{align}
We then set $\Delta^{-1}_{ab}(x,y)$ back to its value that is given in equation (\ref{delta}). 
The first term in equation (\ref{prf}) is consistent with $\Gamma[\alpha]$ having the form
\begin{align}\label{gamma10}
\Gamma[\alpha]=\int d^4xd^4y ~\frac{1}{2}\alpha^a(x)\Delta^{-1}_{ab}(x,y)\alpha^b(y)+\Gamma_{\rm int}[\alpha]
\end{align}
Since $\alpha$ is to be determined so that the one-point function of $\tilde A_a(x)$ vanishes, 
the two-point function for $\tilde A_a(x)$ which occurs in the second term on the right-hand-side of (\ref{prf}) must be connected. Thus, 
$\left. \frac{\delta}{\delta \Delta^{ab}(x,y)}\Gamma_{\rm int}[\alpha]\right|_{\alpha} $ is connected.
It is the sum of terms, each term being the result of the removal of a different propagator, $\Delta_{ab}(x,y)$,  from $\Gamma_{\rm int}[\alpha]$. The sum can only be connected if each term in the sum is connected.  Otherwise, it would be a sum of disconnected and connected terms, and therefore 
$\frac{ \int [d\tilde A] e^{-S_{\rm eff}[\tilde A,\alpha]}\tilde A_c(z)\tilde A_d(z') }  { \int [d\tilde A] e^{-S_{\rm eff}[\tilde A,\alpha]}} $
would be a sum of connected and disconnected parts and $\tilde A_a(x)$ would have a non-zero one-point function, which is a contradiction. If every possible removal of a propagator from $\Gamma_{\rm int}[\alpha]$ results in a connected functional, $\Gamma_{\rm int}[\alpha]$ itself must be irreducible.

Since $\Gamma_{\rm int}[\alpha]$ is an irreducible functional, the terms in its functional Taylor expansion must also be irreducible, they are ($-1$ times)  the irreducible defect current correlation functions which are obtained by sewing together the one-loop correlation functions $\tilde \Gamma^{a_1\ldots a_n}(x_1,...,x_n)$
for $n\geq 3$, which can be visualized as an infinite tower of vertices in a non-local field theory, using the defect-to-defect propagator which is obtained by taking the inverse of the bulk  differential operator
$$
\Delta^{-1}_{ab}(x,x')+\delta(z)\delta(z')\Gamma_{ab}(x,x')
$$
and setting the endpoints on the defect.  This results in the full irreducible  functions $  \Gamma^{a_1\ldots a_n}(x_1,...,x_n)$ which appear in equation (\ref{gamma1}), all of whose
external points are on the defect.   The full effective action therefore necessarily has the form given in equation (\ref{gamma1}) which is a defect field theory.

In this section, we have demonstrated that the full effective has the form (\ref{gamma1}).  The defect quantum field theory remains a defect quantum field theory after radiative corrections are included.  Indeed, this implies that the equation of motion for the expectation value of the vector potential, and therefore the electric and magnetic fields, is that of a free photon in the bulk coupled to correlated defect currents.   In the next sections we will study the properties of the solution of the quantum corrected Maxwell's equations in the weak field regime and in a few different scenarios.

  \section{Quantum corrected Maxwell Equations in the linear regime}
  
The quantum corrected Maxwell equations are the equation of motion (\ref{max}) for the effective action (\ref{gamma1}).  In the regime of linear electrodynamics, where the terms that are of higher order than two in the gauge fields are negligible, either because the sources which induce the gauge fields are small or the $n$-point functions themselves are small, we can simply focus on the quadratic term in the effective action, the irreducible defect current-current correlation function,
\begin{align*}
\frac{1}{2}\int d^3xd^3y\Gamma^{ab}(x,y)\alpha_{a}(x)\alpha_{b}(y)
,~
\Gamma^{ab}(x,y) = \int \frac{d^3k}{(2\pi)^3}e^{ik\cdot(x-y)}\Gamma^{ab}(k),~k^a=(k^0,k^x,k^y)
\end{align*} 
Here, we have assumed translation invariance for displacements along the defect worldvolume. We will also assume that the Ward-Takahashi identity,
$k_{a}\Gamma^{ab}(k)=0 = k_b\Gamma^{ab}(k)$, holds.  

Then we linearize the field equation (\ref{max}) and we find the linearized equation
\begin{align}\label{max_1}
&\left[-(\partial_z^2+\partial^c\partial_c)\delta^{ab}+\partial^a\partial^b\right]\alpha_b(x) +\delta(z)
\int d^4x' \delta(z')\tilde\Gamma^{ab}(x,x') \alpha_b(x')=j^a(x)
\end{align}
This equation has the solution
\begin{align}
 &\alpha^z(k,z)=0 \label{solution_22}\\
&\alpha^a(k,z) =  \int dz'  \frac{e^{-k|z-z'|}}{2k}\left(\delta^a_{b}-\frac{k^ak_b}{k^2}\right)j^b(k,z')  - \int dz' \frac{1}{2} |z-z'| \frac{k^ak_b}{k^2} j^b(k,z') \nonumber \\
&~~~~~~~~~~~~~~~~~~~~~~~~~~~~~~~~
-  \int dz'  \frac{e^{-k(|z|+|z'|)}}{2k}\left[\frac{ \Gamma}{2k+\Gamma}\right]^a_{~b} \left(\delta^b_{c}-\frac{k^bk_c}{k^2}\right)j^c(k,z')
 \label{solution_1}
 \end{align}
 where we have assumed that the current density is Fourier transformable in the $(t,x,y)$ variables and
 we define the Fourier transforms over $(t,x,y)$ as
\begin{align}
&\alpha_\mu(x_a,z) = \int \frac{d^3k}{(2\pi)^3}e^{ik_b(x-y)^b}\alpha_\mu(k,z),~j_\mu(x_a,z) = \int \frac{d^3k}{(2\pi)^3}e^{ik_b(x-y)^b}j_\mu(k,z)
\label{partial_fourier_transforms}\\
& k^b=(k^0,k^x,k^y)
,~k=\sqrt{k_x^2+k_y^2+k_0^2}
\nonumber
\end{align}
Equation (\ref{solution_22}) reminds us that we are in the $\alpha^z=0$ gauge.  
The first two terms on the right-hand-side of equation (\ref{solution_1}) are the solutions of vacuum Maxwell's equations in this gauge with the source $j^a(k,z)$.  We should remember that the source current is conserved, $ik_aj^a(k,z)+\partial_z j^z(k,z)=0$ and that the $z$-component of $j^\mu(x)$ is related to the other components by this continuity equation.  The third term on the right-hand-side of equation (\ref{solution_1}) contains the response of the defect to the induced field.  We shall have much more to say about this response term in the following sections. Finally, 
it is easy to see that if the test current is located on the defect, $j^a(k,z)=j^a(k)\delta(z)$ with $k_aj^a(k)=0$, so that
 \begin{align}
F^{az}(k,z\to 0^+) -F^{az}(k,z\to 0^-) =  j^a(k) -~\left[\frac{\Gamma}{2k+\Gamma}\right]^a_{~b} j^b(k)
 \end{align}
 which contains the boundary current, the first term on the right-hand-side and the induced current, or image current, which is the second
 term on the right-hand-side. In this sense, 
 the fields obey the defect boundary condition (\ref{defect_bc}). 

\section{Defect Conformal Field Theory}

Conformal symmetry restricts the form of the defect correlation functions of operators with definite scaling dimensions.  The defect current $J^a(x)=-e\bar\psi(x)\gamma^a\psi(x)$ is conserved and thus it does not get an anomalous dimension.\footnote{Some more general quantum field theories where it might get an anomalous dimension are discussed in recent interesting work in references \cite{Basa:2019ywr}
-\cite{LaNave:2019mwv}.}.  Its two-point function is then determined by dimensional analysis, Lorentz invariance, and the Ward-Takahashi identity and it contains two parameters, $\chi$ and $\kappa$ in
\begin{align*}
\Gamma_{ab}(k) =\chi \sqrt{k^ck_c}\left(\delta_{ab}-\frac{k_ak_b}{k^2}\right) + \kappa \epsilon_{abc}k^c
\end{align*}
Here, $\kappa $ is a parity and time-reversal symmetry violating parameter which includes the tree-level $\tilde\kappa$ from equation (\ref{lagrangian3}) as well as quantum corrections to it.  If time reversal and parity symmetries are present, $\kappa$ would vanish. On the other hand, $\kappa$ is well known to get contributions from integrating out Fermions with time reversal violating mass terms \cite{Niemi:1983rq}-\cite{Redlich:1983kn}.  Note that on an open, infinite space-time, $\kappa$ is not necessarily quantized and, in a theory with no charge gap, such as the defect conformal field theory, it can obtain nontrivial and non-quantized quantum corrections beyond one-loop order \cite{Semenoff:1988ep}. The parameter $\chi$ is generically non-zero for any system with mobile charged matter on the defect. For example, in the one-loop approximation of the theory with $N$ species of massless defect fermions, it is given by
\begin{align*}
\chi =\frac{Ne^2}{8}+\ldots
\end{align*}
where the ellipses stand for terms of order $e^4$ and higher.  This correction 
can be significant, particularly if $N$ is large.
The quantity which enters the solution of the Maxwell equation, $\left[ \frac{\Gamma}{2k+\Gamma}\right]$, depends on these parameters
and, explicitly, it is 
\begin{align}
\left[ \frac{\Gamma}{2k+\Gamma}\right]_a^{~b}  
= \frac{ \chi(2+\chi)+\kappa^2 }{ (2+\chi)^2+\kappa^2 } \left(\delta_{a}^{~b}-\frac{k_ak^b}{k^2}\right) + \frac{ 2   \kappa}{(2+\chi)^2+\kappa^2} 
\frac{ \epsilon_a^{~bc}k_c } {k}
\label{gamma}
\end{align}

The two terms in the above equation are the time reversal non-violating and the time reversal violating response of the defect.
When they are plugged in to the linearized Maxwell equations, they both have interesting consequences.  In the following sections, we will
examine some of the possibilities.

\section{Defect conformal field theory with a Static Point Test Charge}

Let us consider the response of the defect conformal field theory to the presence of a static point charge $Q$ located at the space-point $(x_0,y_0,z_0)$, that is the current density
\begin{align*}
j^\mu(x)=Q\delta^\mu_0\delta(x-x_0)\delta(y-y_0)\delta(z-z_0)
\end{align*}
If we plug  this current density and the response function given in equation (\ref{gamma}) into the general solution given in equation (\ref{solution_1}), we can find the classical photon field that is induced by a static point charge located at the space-point $(x_0,y_0,z_0)$.  The vector potential is
\begin{align}
\alpha_0(x)=&\frac{Q}{4\pi}\left\{\frac{1}{\sqrt{ (x-x_0)^2+ (y-x_0)^2+(y-z_0)^2}}- \right. \nonumber \\
&\left. ~~~~~~~~~~~ - \frac{\chi(2+\chi)+\kappa^2}{(2+\chi)^2+\kappa^2} \frac{1}{\sqrt{(x-x_0)^2+ (y-x_0)^2+(|z|+|z_0|)^2}} \right\} \label{alphazero}\\
\alpha_x(x)=&\frac{g}{4\pi} \frac{y-y_0}{(x-x_0)^2+(y-y_0)^2}\left(1-\frac{|z|+|z_0|}{\sqrt{(x-x_0)^2+(y-y_0)^2+(|z|+|z_0|)^2}}\right) \label{alphax}\\
\alpha_y(x)=& -\frac{g}{4\pi} \frac{x-x_0}{(x-x_0)^2+(y-y_0)^2}\left(1-\frac{|z|+|z_0|}{\sqrt{(x-x_0)^2+(y-y_0)^2+(|z|+|z_0|)^2}}  \right) \label{alphay}\\
\alpha_z(x)=&0 \label{alphaz} \\
g=&Q~\frac{ 2\kappa}{(2+\chi)^2+\kappa^2} \label{monopolecharge}
\end{align}
Here, we have Wick rotated the result back to Minkowski space so that the field strength tensor that is gotten from this vector potential represent physical electric and magnetic fields.\footnote{This rotation back to Minkowski space avoids some spurious factors of $i=\sqrt{-1}$ associated with time reversal violation and Euclidean space. }

The temporal component of the gauge field, $\alpha_0(x)$, in equation (\ref{alphazero})  has two terms.   The first term is the  Coulomb potential of the point charge  $Q$ which is located at $(x_0,y_0,z_0)$.  The second term is
the Coulomb potential of an image charge $-\frac{\chi(2+\chi)+\kappa^2}{(2+\chi)^2+\kappa^2}Q$ which is located at position $(x_0,y_0, -|z_0|{\rm sign}(z))$.   
The $z$-coordinate, $-|z_0|{\rm sign}(z)$, always has the opposite sign to the observer value of $z$. Consequently,  it is always located on the side of the defect that is opposite to the position of the observer, which is located at $(x,y,z)$. It can therefore be either on top of the test charge, if the observer and test charge are on opposite sides of the defect, or at the mirror image point relative to the test charge, if the observer is located on the same side of the defect as the test charge.

If the observer, at point $(x,y,z)$,  is on the same side of the defect as the point test charge (${\rm sign}(z)={\rm sign}(z_0)$,  
 the Coulomb field is 
\begin{align*}
\alpha_0=&\frac{Q}{4\pi}\left[\frac{1}{\sqrt{ (x-x_0)^2+ (y-y_0)^2+(z-z_0)^2}}- 
\frac{  \frac{\chi(2+\chi)+\kappa^2}{(2+\chi)^2+\kappa^2}}{\sqrt{(x-x_0)^2+ (y-y_0)^2+(z+z_0)^2}} \right] \\
\approx&  \frac{Q}{4\pi}\frac{2 (2+\chi)}{ (2+\chi)^2+\kappa^2 }\frac{1}{\sqrt{ (x-x_0)^2+ (y-y_0)^2+(z-z_0)^2}} ~,~~|z|>>|z_0|
\end{align*}
 If the observer and the test charge are on opposite sides of the defect, the image and test charge are at the same position.
In that case the Coulomb field is still that of a point charge with value
\begin{align}\label{opposite}
\alpha_0(x,y,z)=&  \frac{Q}{4\pi}~\frac{ 2(2+\chi)}{ (2+\chi)^2+\kappa^2 }\frac{1}{\sqrt{ (x-x_0)^2+ (y-y_0)^2+(z-z_0)^2}}
\end{align}
In both cases, when seen from large distances, the charge is screened from $Q$ to $ \frac{ 2(2+\chi)}{ (2+\chi)^2+\kappa^2 }Q$ which, since $\chi$ is generically positive, has magnitude less than $Q$.  This screening can be significant. It is present when there is time reversal symmetry, that is, when $\kappa=0$,  and it is enhanced  when time reversal symmetry is absent and  $\kappa\neq 0$.  
We note that this screening of $Q$ is a  dielectric effect which is present even when there is no dielectric in the bulk.  In particular, if the defect lies between the observer and the test charge, equation (\ref{opposite}) tells us that, no mater what the position and orientation of the defect, the Coulomb field is completely isotropic and the net effect is to attenuate the magnitude of the charge. {\it  It is as if the bulk were filled with an isotropic dielectric material, even when no dielectric is present. }

\begin{figure}[h!]
\centerline{\includegraphics[scale=0.5]{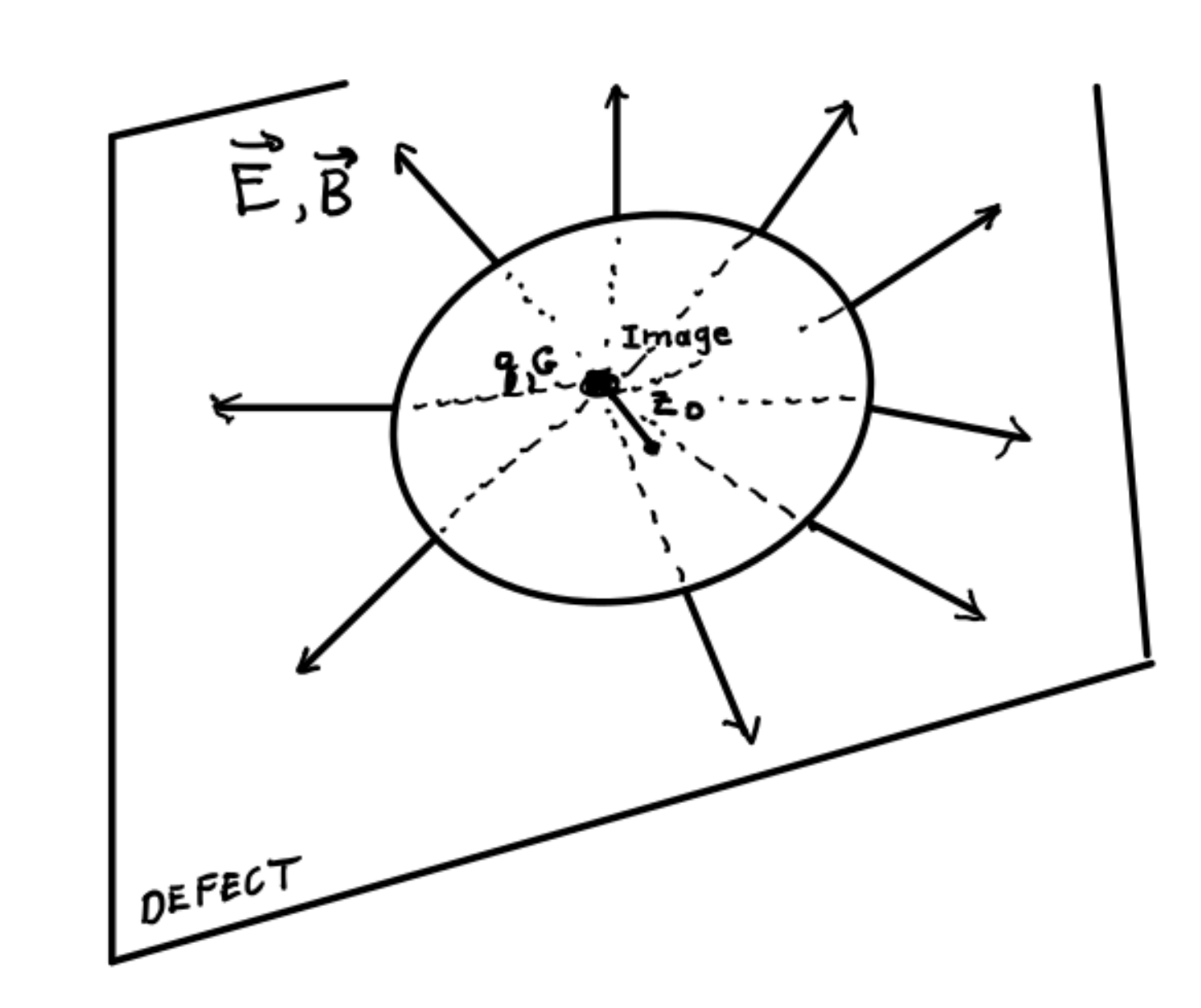}}
\caption{\small Lines of electric and magnetic flux apparently emanating from the image electric and magnetic monopole which 
always appears on the side of the defect that is opposite to the observer.   }
\label{figure1}
\end{figure}

 In the time reversal violating system, when $\kappa$ is nonzero, the spatial components of the vector potential in equations (\ref{alphax}) and (\ref{alphay}) are non-zero and they coincide with the vector potential of a Dirac magnetic monopole in the $a_z=0$ gauge.  The magnetic field, $\vec B=\vec\nabla\times\vec\alpha$, is radially symmetric, 
\begin{align} 
\left( B_x,B_y,B_z\right) = ~\frac{g}{4\pi}~ \frac{\left( x-x_0,y-y_0,z+|z_0|
{\rm sign}(z)\right)}{ |\vec x-\vec{\tilde  x}_0|^3}    {\rm sign}(z)
\end{align}
where $\vec{\tilde x}_0=\left( x_0,y_0,-|z_0|
{\rm sign}(z)\right)$ is the location of the image charge. The monopole is depicted in firgure \ref{figure1}. 
The monopole charge seen by an observer at $(x,y,z)$ is 
\begin{align}
G=\frac{Q}{2}~\frac{2 \kappa}{(2+\chi)^2+\kappa^2}{\rm sign}(z)
\end{align}
and the $z$-component of its position is at $|z_0|{\rm sign}(z)$.  

This is a peculiar monopole in that the magnetic charge has different signs and the monopole has different positions when it is viewed from
different sides of the defect.  Physically, it is not a monopole at all, but a bundle of flux tubes which emulates a monopole field when it is viewed only from one or the other side of the defect.  Since the total magnetic monopole moment, taking into account both sides, is zero, one might expect that the magnetic charge need not be quantized.    This non-quantization of the magnetic charge is  also consistent with the observer always seeing both the monopole singularity and the Dirac string as being in the inaccessible region on the opposite side of the defect.  Note that the monopole charge does not obey the Dirac quantization condition even when the Chern-Simons level is quantized, that is, when $\kappa=\frac{e^2}{2\pi}\cdot$integer, even if $\chi=0$.

 Synthetic or emergent monopoles have previously been found in several contexts, spin ice \cite{spin_ice_1}-\cite{spin_ice_4}, cold atoms \cite{cold_atoms_1}-\cite{cold_atoms_2} and as image charges in certain topological insulators \cite{tme}.  The monopole field profile of image charges in the topological insulating Hall states of suspended graphene, the subject of our next subsection, 
  have recently been looked for \cite{tme_1}-\cite{tme_2} and perhaps seen \cite{tme_3} in precision experiments. 
 The monopole image charge that we find here is very similar to the one that was discussed for the topological insulator.     We will study the quantum Hall  state of graphene in a later section.

 \section{Charge Neutral Graphene}
 
  Graphene is a one-atom thick layer of carbon atoms  where electrons within $\sim 1ev$ of the Fermi energy obey an emergent massless Dirac equation and have a linear dispersion relation $\omega (k)= v_F | k|$ with Fermi velocity $v_F \sim c/300$. It has been studied as an analog of relativistic field theory \cite{SEM} where relativistic quantum mechanics and field theory phenomena special to 2+1 space-time dimensions could be realized in nature.  
When it is coupled to a dynamical photon, which resides in the bulk surrounding the graphene layer, the composite system bears a close resemblance to the defect quantum field theories that we have been discussing so far.   However, since the vacuum speed of light, $c$, which we have set equal to one in this paper, and the 
 speed of the emergent massless graphene electron are different, graphene does not exhibit Lorentz invariance or conformal 
 invariance.   Moreover, in the perturbative regime, the electron speed has a nonzero beta function and it is not scale invariant. 
 In spite of this, graphene does have approximate scale invariant behaviour over a large dynamical range.  In particular, some of its
 electromagnetic properties, such as its AC conductivity, are to a good approximation frequency independent and scale invariant   from milli-volt to electron volt energy scales \cite{NOV} \cite {KIM} \cite{WU}-\cite{Juricic:2010dm}.  What is more, its value is very 
 close to the value given by a computation with free fermions.  If we postulate that the Fourier components of the charge and current
 densities are in this regime of approximate scale invariance, even without full conformal invariance, graphene should exhibit screening
 behaviour similar to what we found for a defect conformal field theory.  We would expect that the defect correlation function has
 the form
 \begin{align}
 \Gamma_{00}(q) = \frac{e^2}{2}\frac{1}{v_F}\sqrt{\vec q^2}+\ldots
 \end{align}
 where $v_F$ is the electron fermi-velocity and we have put $N=4$ to account for the spin and valley degeneracy of graphene. 
The response to the presence of a test charge can be extracted from equation (\ref{alphazero}) as 
 \begin{align}
\alpha^0(k,z) =  \frac{Q}{4\pi} &\left\{ \frac{1}{\sqrt{(x-x_0)^2+(y-y_0)^2+(z-z_0)^2}} - \right. \nonumber \\
&\left. -\frac{e^2/v_F}{e^2/v_F+4} \frac{1}{\sqrt{(x-x_0)^2+(y-y_0)^2+(z-|z_0|{\rm sign(z)})^2}} \right\}
  \label{solution_3}
 \end{align}
 This screening effect is very small.  In the natural units in which we are working, $e^2\approx 4\pi/137$ and with the  graphene
 Fermi velocity is $v_F\approx 1/300$  
 $$
 \frac{e^2/v_F}{e^2/v_F+4}\approx  0.87
 $$
 and test charges should be highly screened
 $$
 Q~\to~0.13Q
 $$  
 This should be a testable effect. As in all of the previous cases, 
 if the graphene screen is located between the observer and the charge, the electric field is still that of a point charge, the screening effect is completely isotropic, but the value of the charge is reduced by ninety percent.   If the observer and the charge are on the same side of the screen, the screening of the electric monopole charge by ninety percent is still seen from a large distance, but closer up the charge and its image have electric dipole and higher
 moments. 
 
\section{Quantum Hall Phase}

The analysis of the previous parts of this section can be applied  to layer of material with a controllable conduction electron density such as graphene which is tuned so that it is in a quantum Hall  phase. The quantum Hall phase is the state of the system which occurs at low temperature and in a strong magnetic field, where, as a function of the ratio of density to magnetic field, the material exhibits plateaux of quantized Hall conductivity.  The system is in a ``quantum Hall phase'' when its parameters are tuned so that it is on such a Hall conductivity plateau.  
It has a non-zero Hall conductance, $\sigma_{xy}$ and it has vanishing longitudinal conductivity $\sigma_{xx}=0$.  The modification of our formalism which would be needed would introduce an additional constant background magnetic field $B_z$.  This is straightforward and it could be done by adding a large solenoid to the external current $j^\mu(x)$.  We will not spell out the details here.  We do note that our discussion does not depend on Lorentz invariance, or conformal invariance of the underlying system.  It is enough that its phenomenology is described at small momenta by the Chern-Simons effective action.   In our language, the quantum Hall state has 
the parameter $\chi$ vanishing and $\kappa = \sigma_{xy}$.   With our units where $\hbar=1$ and $c=1$, and in an integer quantum Hall state, $\sigma_{xy}=\frac{n}{2\pi}e^2$ where $n$ is an integer.  In the case of the anomalous integer quantum Hall effect in graphene \cite{NOV} \cite{KIM}, $n$ would be replaced by $4n+2$ where $n$ is an integer.   Such states in graphene are very pronounced, in fact they are even visible at room temperature. 

 Then, in the presence of a static point charge $Q$ located at the space-point $(x_0,y_0,z_0)$, the vector potential is obtained by setting $\chi$ to zero and $\kappa$ to $ \sigma_{xy}$ in equations (\ref{alphazero})-(\ref{alphaz}).  Let us assume that $z$ and $z_0$ have opposite signs, that is, that the graphene ``screen'' is located between the observer and the test charge.  Then, the test charge appears to the observer as a dyon, a particle with both electric and magnetic charge, in this case, it has an electric monopole charge,  which is reduced by screening to the value
\begin{align} 
  q=\frac{Q}{1+\sigma_{xy}^2/4}   
\end{align}
and it has a magnetic monopole charge
\begin{align}
g=\frac{  2Q\sigma_{xy}}{4+\sigma_{xy}^2} {\rm sign}(z)
\end{align}  The presence of the magnetic charge is very similar to the one that is predicted to appear in some topological insulators \cite{tme}.  In fact, the layer in a Hall state is an example of a topological insulator.  This monopole image charge has   been looked for in experiments  \cite{tme_1}- \cite{tme_3}.  
The remarkable dielectric effect, the suppression of the electric charge, and its isotropy, particularly the fact that the isotropy is independent of the
orientation or position of the defect, as long as the defect is between the test charge and the observer, could also be sufficiently dramatic to be observable.
It could also have applications, for example, it one wanted to design a device which had a variable and isotropic dielectric constant.  Here, the dielectric
constant is $\varepsilon=1+\sigma_{xy}^2/4$ and, by changing Hall plateaus, that is, changing the integer in $\sigma_{xy}=\frac{n}{2\pi}e^2$ one could change the dielectric constant.

  \section{Metallic defects}

It is also interesting to apply our formalism to a metallic state of a defect.   This state has a finite density of of mobile charges and it has no charge gap.
It is neither scale invariant nor Lorentz invariant.  
We shall assume that  gauge, rotation and translation symmetries remain intact. Then $\Gamma(k)$ is a matrix which it is convenient to write as a sum of
 two projection operators, 
\begin{align*}
\Gamma_{ab}(k)= \chi(k) \Pi_{ab}^E(k)+ \xi(k) \Pi_{ab}^M(k) 
\end{align*}
where $\chi(k_0,k)$ and $\xi(k_0,k)$ are two functions of the frequency, $k_0$ and the modulus of the wave-vector $k=\sqrt{\vec k^2}$. Here, we will consider only a time-reversal invariant metal.  The electric and magnetic projection operators, $\Pi_{ab}^E(k)$and $ \Pi_{ab}^M(k)$, respectively, have the properties 
\begin{align}
&\Pi^E_{ab}=\Pi^E_{ba}~,~~\Pi^M_{ab}=\Pi^M_{ba}\\
&k^a\Pi^E_{ab}(k)=0~,~~k^a\Pi^M_{ab}(k)=0\\
&\Pi^E_{ab}+\Pi^M_{ab}= (k_0^2+\vec k^2)\delta_{ab}-k_ak_b\\
&(\Pi^E)^2(k)=(k_0^2+\vec k^2)\Pi^E(k)\\
&(\Pi^M)^2(k)=\vec k^2\Pi^M(k)
\end{align}
and they  have the explicit forms 
\begin{align*}
\Pi^E =  
\left[ \begin{matrix} \vec k^2 & -k_0k_j \cr -k_0 k_i & 
k_0^2\delta_{ij} \cr\end{matrix}\right]
~,~~
\Pi^M = 
\left[ \begin{matrix} 0 &0 \cr0 &  \vec k^2 \delta_{ij} -   k_ik_j  \cr\end{matrix}\right]
\end{align*}
These appear in the electric and magnetic parts of the effective action which, using
translation, rotation and gauge invariance, can be seen to have the form
\begin{align*}
S_{\rm effective}=\frac{1}{2}\int d^3xd^3y \left[\vec  E(x)\chi(x-y)\cdot \vec E(y)+B(x)\xi(x-y)B(y)  
  +\ldots\right]
\end{align*}
where $\vec E=\partial_0\vec\alpha -\vec \partial \alpha_0$ and $B=\partial_x\alpha_y-\partial_y\alpha_x$ are the electric and magnetic
fields on the defect and $\chi(x-y)$, $\xi(x-y)$ are the Fourier transforms of $\chi(k_0,k)$ and $\xi(k_0,k)$. When the system is
Lorentz invariant, $\chi(k_0,k)=\xi(k_0,k)$ and both are functions only of $k_ak^a=k_0^2+\vec k^2$. This  is
 so in the conformal field theory which we studied in earlier sections.   

Our discussion in this section will depend only on the generic form of the defect current-current correlation function, $-\Gamma_{ab}(k_0,k)$. 
Of course, unlike in the case of a conformal field theory, this quantity is generically a complicated function of frequency and wave-vector. 
There is a limit of our discussion where the results only depend on the small frequency and wave-number limits.  In those limits, 
the correlations functions have a certain generic form which can be described by a few parameters,
\begin{align}\label{matal_1}
\Gamma_{00}(k_0,  k)=\alpha_1\mu +{\alpha_2}\frac{ k^2}{ \sqrt{  k^2+k_0^2}}+\ldots
\end{align}
where $\alpha_1$ and $\alpha_2$ are dimensionless numbers which characterize the metallic state of the defect, 
$\mu$ is the chemical potential and the ellipses denote higher orders in powers of $k_0$ and $k$. 
An explicit one-loop computation of $\Gamma_{ab}(k_0,k)$ for massless defect Fermions with finite carrier density which is induced by 
the presence of a chemical potential $\mu$ is reviewed in the Appendices.   It yields approximate values of the parameters $\alpha_1$ and $\alpha_2$ as
\begin{align}
\alpha_1 =\frac{e^2N}{2\pi} +\ldots~,~\alpha_2=\frac{e^2N}{16}+\ldots
\end{align}
where corrections would be of higher orders in the coupling constant $e^2$. Moreover, in the region where $k\leq2\mu$,  the leading order in $e^2$ expression $\Gamma_{00}= \frac{e^2N}{2\pi} \mu+\frac{e^2N}{16} k$ is exact to all orders in $k$.

If we consider a static point charge, where $j^0=Q\delta(x-x_0)\delta(y-y_0)\delta(z-z_0)$, plugging $\Gamma_{00}$ into equation (\ref{solution_1}) leads to
  \begin{align}
&\alpha^0(\vec x,z) =  \int \frac{d^2k}{(2\pi)^2} \left\{ \frac{e^{-k|z-z_0|}}{2k}  
-  \frac{e^{-k(|z|+|z_0|)}}{2k}\left[\frac{1 +\frac{\alpha_2}{\alpha_1\mu}k+\ldots}{1 +\frac{(2+{\alpha_2})}{\alpha_1\mu}k 
+\ldots}\right]
\right\}Qe^{ik\cdot(x- x_0)}  \nonumber \\
&=\frac{Q}{4\pi}\left\{\frac{1}{\sqrt{(x-x_0)^2+(y-y_0)^2+(z-z_0)^2}} - \frac{1}{\sqrt{(x-x_0)^2+(y-y_0)^2+(z+|z_0|{\rm sign}(z))^2}}\right. 
\nonumber\\
&\left. ~~~~~~~~~~~~~~~~~~~~~~~~~~~+\frac{2}{\alpha_1\mu} \frac{1}{ (x-x_0)^2+(y-y_0)^2+(z+|z_0|{\rm sign}(z))^2 }+\ldots \right\}
 \label{metal} \end{align}
 where the ellipses stand for corrections which are suppressed by higher orders in factors of $\frac{1}{\mu\cdot{\rm distance}}$. 
We see that, particularly when $|z_0|{\rm sign}(z)=-z_0$, that is, when the observer and the test charge are on opposite sides of the defect, the
electric monopole terms, the first two on the right-hand-side of equation (\ref{metal}) cancel exactly.  The electric monopole moment is completely
screened.  It is replaced by the term in the second line of equation (\ref{metal}) which falls off like the inverse of distance squared, rather than distance, from the image charge. Generically, there would be corrections to this formula with higher powers of the inverse distance. 
We point out that, in this way, the metallic defect differs from a sheet of metal with finite thickness where one would expect that the electric field on the side which is opposite to the test charge is identically zero. 

Some intuition for this result can be elicited by recalling the ``monolayer'' nature of the defect.  If we consider a point charge sitting on the defect itself, the lines of electric flux emanate from the point charge directly to the bulk.  The effect of the metallic defect is to make those lines normal to the defect.  This essentially makes the monopole charge appear to be a dipole.  In fact, a careful study of how charges in the metallic defect interact with each other sees that they interact like dipoles, rather than charge monopoles.  This is strikingly different from what happens in a bulk metal where any electric field must decay exponentially with distance and all electrostatic interactions become short ranged with exponential tails.   
   
 \section{Summary}
 
 We have examined the electromagnetic properties of a defect quantum field theory where the interacting charged particles reside on a 2+1-dimensional defect and the surrounding 3+1 dimensional bulk is occupied by a free photon field.  We showed that the fully quantum corrected quantum theory remains a defect field theory in that the classical equations of motion for the electromagnetic fields which are induced by the presence of   test charge and current densities are given by field equations where all of the interaction terms, including nonlinearities as well as corrections to the linear terms,  are confined to the defect.  
 
 We then found a solution of the linearized Maxwell equations and we discussed the properties of the solution.  The two most striking properties are the screening of charge and the fact that, for a defect which violates time reversal symmetry, point charges are screened and they appear to have a magnetic monopole charge. This effect is particularly simple when the test charge and the observer are on opposite sides of the defect.  Then the position of the image charge coincides with the position of the test charge, and the test charge simply appears with a reduced electric charge and possibly a magnetic charge with the fully isotropic electric and magnetic fields of a point source.   When the test charge and the observer are on the same side of the defect, the image is located at the point that is mirror image through the defect and the electric and magnetic fields are a superposition of the fields of two point sources.  The long ranged part, the electric monopole moment is still screened as before, and the magnetic monopole moment is totally isotropic, since only the image has magnetic charge. 
 
In a conformal defect field theory this screening effect depends on two parameters, one time reversal symmetric and one time reversal violating. 
This includes the quantum Hall state of a two-dimensional defect. In that case, the defect is described by a topological field theory which for our purposes can be regarded as a limit of a conformal field theory where longitudinal charge transport is suppressed.   It is also related to other topological insulators and the fact that the induced charge appears as a magnetic monopole has been posited already ten years ago and it has recently been searched for in exeriments.  

We have also discussed the effect in some cases which are neither conformal nor topological.  If the defect is a metal the inverse distance dependence of the Coulomb field is replaced by an inverse distance squared behaviour.  This differs from what happens in a bulk metal where the electric potential  would off exponentially. It also differs from a thick sheet of metal which would screen screen an electric field completely.  This is particularly interesting for the case where the observer and test charge are on opposite sides of the defect.  If they are on the same side, the point charge is replaced by a dipole consisting of the point charge and its image, located at its mirror image point.

 \appendix

 \section{Appendix: The Insulator and the Semi-metal: One-loop computations}
 
 In these appendices, we will summarize some computations of the time-ordered Euclidean space current-current correlations functions in the one-loop approximation for examples of the physical states of the defect field theory that we are interested in.  These computations are not original and the reader should be able to find them elsewhere in the literature.  We summarize them here for the benefit of the reader. At risk of being overly pedantic, we provide many details of the computations, also to help the reader who might want to modify or generalize the arguments of this paper. 
 
 We shall begin in this first appendix with a computation of the current-current correlation function of an insulator and a semi-metal.  We will model the insulator by the vacuum state of massive Dirac Fermions where the mass is time-reversal invariant.  We will model the semimetal by taking the massless limit at zero charge density.  The semimetal is our example of a conformal field theory when the speed of the Dirac fermion is equal to the speed of the photon in the bulk, and it emulates graphene with a Coulomb interaction, which is not a conformal or Lorentz invariant model when the two speeds are different.   In both cases, our computation of the defect current-current correlation function is at one loop order.   We begin with the Euclidean momentum space Feynman integral
 $$
 \Gamma_{ab} (q)= e^2\int \frac{d^{2\omega}p}{(2\pi)^{2\omega}}{\rm Tr} \gamma_a\frac{1}{[\slashed p-iM]}\gamma_b\frac{1}{[\slashed p+\slashed q-iM]} $$ 
 where the trace is over the three $2\times2$ Euclidean $\gamma$-matrices, which we can take as the Pauli matrices,  and over the $N$ species of fermions.  We will perform the computation in Euclidean space.  We will consider only  time-reversal invariant systems here.    
 In order to maintain time-reversal invariance, we shall assume    that the Fermion mass matrix $M$ is a traceless $N\times N$ matrix obeying ${\rm Tr}M=0$ and $M^2=m^2$, the square of the fermion mass.  In this case, the parity anomaly cancels.  Then, 
 taking the traces\footnote{We assume here that the dimension of the Dirac matrices is $2\times2$.  We could be more general at this point, but it will not be needed at one loop order since the ultraviolet divergences will cancel and the integral is finite.}, and introducing Feynman parameters and dimensional regularization, we get 
 $$
 \Gamma_{ab} (q)= 2Ne^2\int_0^1d\alpha \int \frac{d^{2\omega}p}{(2\pi)^{2\omega}} \frac{ 
  -2\alpha(1-\alpha)q_aq_b+\delta_{ab}[(\frac{1}{\omega}-1)p^2+\alpha(1-\alpha)q^2-m^2]
}
 {[p^2+\alpha(1-\alpha)q^2+m^2]^2}
 $$ 
 where we have used the symmetry of the integrand to drop terms that are odd in $p$ and substitute $p_ap_b$ by $\frac{1}{2\omega}p^2\delta_{ab}$.
 Then, we can use standard dimensional regularization formulae to do the integration over $p$ to get
 $$
 \Gamma_{ab} (q)
 =4Ne^2\frac{\Gamma[2-\omega] }{(4\pi)^\omega}[\delta_{ab}q^2 - q_aq_b]\int_0^1d\alpha \frac{ 
\alpha(1-\alpha)
}
 {[\alpha(1-\alpha)q^2+m^2]^{2-\omega}}
 $$ 
 which is finite in three dimensions, when $\omega=3/2$, 
 $$
  \Gamma_{ab} (q)
=\frac{Ne^2 }{2\pi}[\delta_{ab}q^2 - q_aq_b]\int_0^1d\alpha \frac{ 
\alpha(1-\alpha)
}
 {\sqrt{\alpha(1-\alpha)q^2+m^2}}
 $$ 
  $$
 ~~~~~~~~~~~~~~~~=\frac{Ne^2 }{2\pi q}[\delta_{ab}q^2 - q_aq_b]\left\{\frac{m}{2q} 
 +\left(\frac{1}{4}-\frac{m^2}{q^2}\right)\arctan\frac{q}{2m}
 \right\}
 $$
 $$
 \approx  \frac{Ne^2 }{8 q}[\delta_{ab}q^2 - q_aq_b]~~m\to 0 ~~~{\rm semimetal}
 $$ 
  $$
 \approx \frac{Ne^2 }{12\pi |m|}[\delta_{ab}q^2 - q_aq_b]~~m\to \infty ~~~{\rm insulator}
 $$ 
 
  The result for the semi-metal is the entire contribution for massless Fermions at one-loop order.
    This form of the correlation function, when combined with the Kubo formula, and corrected to replace the vacuum speed of light
  by the speed of the graphene electron, gives an accurate formula for the AC conductivity of graphene in the limit where
  the frequency is much greater than the temperature \cite{MIS}.

  \section{Appendix: Metallic Defect to One Loop Order}
  
  We will model a metal by massless Dirac Fermions at finite density.  We will control the density by introducing a chemical potential.  We will compute all of the components of the Euclidean current-current correlation function and then we will get the Lorentizian functions by analytic continuation.  Here, we will focus on the time reversal invariant case only.
  We begin with the expression
   
 $$
 \Gamma_{ab}(q) = e^2N\int \frac{d^{3}p}{(2\pi)^{3}}{\rm Tr}\frac{ \gamma_a\slashed p\gamma_b(\slashed p+\slashed q)}{p^2(p+q)^2}
 $$
 where the chemical potential is contained in the temporal component of the momenta as $p_0-i\mu$ so that 
$$
 \Gamma_{ab}(q) = 2Ne^2\int \frac{d^{2\omega}p}{(2\pi)^{2\omega}} \int \frac{dp_0}{2\pi}
 \frac{\gamma_{ab}(p_0-i\mu,\vec p)}
 {[(p_0-i\mu)^2+\vec p^2][(p_0+q_0-i\mu)^2+(\vec p+\vec q)^2]}
 $$  
 where 
 $$
 \gamma_{ab}(p_0-i\mu ,\vec p)= 
 p_a(p+q)_b+(p+q)_ap_b-\delta_{ab}p\cdot(p+q)
 $$
 Here, we have dimensionally regulated the integration over spatial momenta.  The time component of the loop momentum is integrated
 along the real axis in the complex $p_0$-plane. 
 The $p_0$ integral is done by completing the contour in the upper half of the complex $p_0$-plane and using Cauchy's theorem:
 $$
  \int \frac{dp_0}{2\pi}
 \frac{\gamma_{ab}(p_0-i\mu,\vec p)}
 {[(p_0-i\mu)^2+\vec p^2][(p_0+q_0-i\mu)^2+(\vec p+\vec q)^2]}
 =\frac{\gamma_{ab}(i|\vec p| ,\vec p)\theta(\mu+|\vec p|)}
 {[2|\vec p|][q_0+i|\vec p| -i|\vec p+\vec q|][ q_0 +i|\vec p|+i|\vec p+\vec q|]}
$$
$$
+ \frac{\gamma_{ab}(-i|\vec p| ,\vec p)\theta(\mu-|\vec p|)}
 {[-2i|\vec p|] [ q_0 -i|\vec p|-i|\vec p+\vec q|][ q_0 -i|\vec p|+i|\vec p+\vec q|]}
 +\frac{\gamma_{ab}(-q_0+i|\vec p+\vec q| ,\vec p)\theta(\mu+|\vec p+\vec q|)}
 {[-q_0 +i|\vec p+\vec q| -i|\vec p|][-q_0+i|\vec p+\vec q|+i|\vec p|][2|\vec p+\vec q|]}
$$
$$
 +\frac{\gamma_{ab}(-q_0 -i|\vec p+\vec q| ,\vec p)\theta(\mu-|\vec p+\vec q|)}
 {[-q_0-i|\vec p+\vec q|-i|\vec p|][-q_0-i|\vec p+\vec q|+i|\vec p|] [-2i|\vec p+\vec q|]}
 $$
 We change variables $\vec p \to -\vec p - \vec q$ in the last two terms (this is a symmetry of the integration over $\vec p$ which we shall
 do later) to present the above expression as 
     $$
 =\frac{ \theta(\mu+|\vec p|) }
 {2|\vec p| }\left[ 
 \frac{\gamma_{ab}(i|\vec p| ,\vec p) }
 { (q_0+i|\vec p|)^2 +(\vec p+\vec q)^2  }
  +\frac{\gamma_{ab}(-q_0+i|\vec p| ,-\vec p-\vec q)}
 { (q_0-i|\vec p|)^2 +(\vec p+\vec q)^2  } \right]
$$
$$
- \frac{ \theta(\mu-|\vec p|)}
 {2|\vec p| }\left[ 
 \frac{\gamma_{ab}(-i|\vec p| ,\vec p) }
 { (q_0-i|\vec p|)^2 +(\vec p+\vec q)^2 }
 +\frac{\gamma_{ab}(-q_0 -i|\vec p| ,-\vec p-\vec q)}
 { (q_0+i|\vec p|)^2 +(\vec p+\vec q)^2 } \right]
 $$
 If we now assume that $\mu>0$ and we subtract the $\mu\to 0$ contribution so that the result is the chemical potential-dependent 
 part of the two-point function which we denote

 $$
 \delta\Gamma_{ab} = Ne^2\int \frac{d^{2}p}{(2\pi)^{2}}\frac{ \theta(\mu-|\vec p|)}
 {|\vec p| }\left[ 
 \frac{\gamma_{ab}(-i|\vec p| ,\vec p) }
 { (q_0-i|\vec p|)^2 +(\vec p+\vec q)^2 }
 +\frac{\gamma_{ab}(-q_0 -i|\vec p| ,-\vec p-\vec q)}
 { (q_0+i|\vec p|)^2 +(\vec p+\vec q)^2 } \right]
 $$
 In components we find 
 \begin{equation}
\delta\Gamma_{00}= -\frac{e^2 N}{4\pi}\int_0^\mu dp\left[\dfrac{\sqrt{\vec{q}^2+\left(q_0-2 i p\right)^2}}{\sqrt{\vec{q}^2+q^2_0}}+\dfrac{\sqrt{\vec{q}^2+\left(q_0+2 i p\right)^2}}{\sqrt{\vec{q}^2+q^2_0}}-2\right]
\end{equation}

\begin{equation}
\delta\Gamma_{0i}= \frac{e^2 N}{4\pi}\dfrac{q_0q_i}{\vec{q}^2}\int_0^\mu dp\left[\dfrac{\sqrt{\vec{q}^2+\left(q_0-2 i p\right)^2}}{\sqrt{\vec{q}^2+q^2_0}}+\dfrac{\sqrt{\vec{q}^2+\left(q_0+2 i p\right)^2}}{\sqrt{\vec{q}^2+q^2_0}}-2\right]
\end{equation}

\begin{align}\label{deltaGammaij}
&\delta\Gamma_{ij}= \frac{e^2 N}{4\pi}\dfrac{q_0^2}{\vec{q}^2}\left(\delta_{ij}-2\dfrac{q_iq_j}{\vec{q}^2}\right)\int_0^\mu dp\left[\dfrac{\sqrt{\vec{q}^2+\left(q_0-2 i p\right)^2}}{\sqrt{\vec{q}^2+q^2_0}}+\dfrac{\sqrt{\vec{q}^2+\left(q_0+2 i p\right)^2}}{\sqrt{\vec{q}^2+q^2_0}}-2\right]
\cr
&+\frac{e^2 N}{\pi}\left(\delta_{ij}-\dfrac{q_iq_j}{\vec{q}^2}\right)\int_0^\mu dp\left[\dfrac{ip(q_0-ip)}{\sqrt{\vec{q}^2+q^2_0}\sqrt{\vec{q}^2+\left(q_0-2 i p\right)^2}}-\dfrac{ip(q_0+ip)}{\sqrt{\vec{q}^2+q^2_0}\sqrt{\vec{q}^2+\left(q_0+2 i p\right)^2}}\right]
\end{align}

The Ward-Takahashi identities 
\begin{equation}
q_0\delta \Gamma_{00}(q)+q_i\delta \Gamma_{0i}(q)=0~,~~~~~q_0\delta \Gamma_{0j}(q)+q_i\delta \Gamma_{ij}(q)=0
\end{equation}
are satisfied by the above expressions. 

\subsection{The functions $\chi(q)$ and $\xi(q)$}

When the defect is not Lorentz invariant but still has space and time translation symmetry the current-current correlation function  $\Gamma_{ab}(k)$ is a matrix
this is determined by two projection operators, 
\begin{align*}
\Gamma_{ab}(k)= \chi(k) \Pi_{ab}^E(k)+ \xi(k) \Pi_{ab}^M(k) 
\end{align*}
When the field theory is Lorentz invariant, $\chi(k)=\xi(k)$ and both of them are functions only of $\vec k^2+k_0^2$.   
Here, the chemical potential violates Lorentz invariance, but not rotation invariance, so these functions depend on the
frequency $k_0$ and the wave-number $k$.

From \eqref{deltaGammaij} one can read   $\delta\chi(q)$  and the $\delta\xi(q)$ which are the chemical potential-dependent parts of 
$\delta\chi(q)$  and  $\delta\xi(q)$
\begin{equation}\label{chi}
\delta\chi(q)=-\frac{e^2 N}{4\pi}\dfrac{1}{\vec{q}^2}\int_0^\mu dp\left[\dfrac{\sqrt{\vec{q}^2+\left(q_0-2 i p\right)^2}}{\sqrt{\vec{q}^2+q^2_0}}+\dfrac{\sqrt{\vec{q}^2+\left(q_0+2 i p\right)^2}}{\sqrt{\vec{q}^2+q^2_0}}-2\right]
\end{equation}
\begin{align}\label{xi}
&\delta\xi(q)=  \frac{e^2 N}{2\pi}\dfrac{q_0^2}{(\vec{q}^2)^2}\int_0^\mu dp\left[\dfrac{\sqrt{\vec{q}^2+\left(q_0-2 i p\right)^2}}{\sqrt{\vec{q}^2+q^2_0}}+\dfrac{\sqrt{\vec{q}^2+\left(q_0+2 i p\right)^2}}{\sqrt{\vec{q}^2+q^2_0}}-2\right]
\cr
&- \frac{e^2 N}{\pi}\dfrac{1}{\vec{q}^2}\int_0^\mu dp\left[\dfrac{ip(q_0-ip)}{\sqrt{\vec{q}^2+q^2_0}\sqrt{\vec{q}^2+\left(q_0-2 i p\right)^2}}-\dfrac{ip(q_0+ip)}{\sqrt{\vec{q}^2+q^2_0}\sqrt{\vec{q}^2+\left(q_0+2 i p\right)^2}}\right]
\end{align}
Upon integration over $p$, $\delta\chi(q)$ and $\delta\xi(q)$ become
\begin{align}\label{integratedchi}
&\delta\chi(q)=\frac{e^2 N  \mu}{2\pi \vec{q}^2}+\frac{e^2 N}{16 \pi \vec{q}^2
   \sqrt{\vec{q}^2+q_0^2}}
\left\{ i q_0 \left(\sqrt{\vec{q}^2+(q_0-2 i \mu )^2}-\sqrt{\vec{q}^2+(q_0+2 i \mu
   )^2}\right)
   \right.\nonumber\\&\left.
  + 2 \mu  \left(\sqrt{\vec{q}^2+(q_0-2 i \mu )^2}+\sqrt{\vec{q}^2+(q_0+2 i
   \mu )^2}\right)
      \right.\nonumber\\&\left.
 + \vec{q}^2 \left[ \tan ^{-1}\left(\frac{2 \mu 
   \left(q_0+\sqrt{\vec{q}^2+(q_0+2 i \mu )^2}\right)}{\vec{q}^2+4 i \mu  (q_0 +
   i \mu)}\right)+ \tan ^{-1}\left(\frac{2 \mu 
   \left(-q_0+\sqrt{\vec{q}^2+(q_0-2 i \mu )^2}\right)}{\vec{q}^2-4 i \mu  (q_0 -i
   \mu)}\right)
       \right.\right.\nonumber\\&\left.\left.
  - i \tanh ^{-1}\left(\frac{q_0 \left(\sqrt{\vec{q}^2+(q_0+2 i
   \mu )^2}-2 i \mu \right)}{\vec{q}^2+q_0 (q_0+4 i \mu )}\right)- i \tanh
   ^{-1}\left(\frac{q_0 \left(2 i \mu -\sqrt{\vec{q}^2+(q_0-2 i \mu
   )^2}\right)}{\vec{q}^2+q_0 (q_0-4 i \mu )}\right)\right]\right\}
   \end{align}

\begin{align}\label{integratedxi}
&\delta\xi(q)=\frac{e^2 N \mu q_0^2 }{2\pi (\vec{q}^2)^2}+\frac{e^2 N \left(2 \vec{q}^2-q_0^2 \sqrt{\vec{q}^2+q_0^2}\right) }{32 \pi  \vec{q}^4 \sqrt{\vec{q}^2+q_0^2}}
\left\{
 i q_0
   \left(\sqrt{\vec{q}^2+(q_0-2 i \mu )^2}-\sqrt{\vec{q}^2+(q_0+2 i \mu )^2}\right)
     \right.\nonumber\\&\left.
   +2
   \mu  \left(\sqrt{\vec{q}^2+(q_0-2 i \mu )^2}+\sqrt{\vec{q}^2+(q_0+2 i \mu
   )^2}\right)
       \right.\nonumber\\&\left.
   -\left(\vec{q}^2+2
   q_0^2\right) \left[ \tan ^{-1}\left(\frac{2 \mu 
   \left(q_0+\sqrt{\vec{q}^2+(q_0+2 i \mu )^2}\right)}{\vec{q}^2+4 i \mu  (q_0 +
   i \mu)}\right)+ \tan ^{-1}\left(\frac{2 \mu 
   \left(-q_0+\sqrt{\vec{q}^2+(q_0-2 i \mu )^2}\right)}{\vec{q}^2-4 i \mu  (q_0 -i
   \mu)}\right)
       \right.\right.\nonumber\\&\left.\left.
  - i \tanh ^{-1}\left(\frac{q_0 \left(\sqrt{\vec{q}^2+(q_0+2 i
   \mu )^2}-2 i \mu \right)}{\vec{q}^2+q_0 (q_0+4 i \mu )}\right)- i \tanh
   ^{-1}\left(\frac{q_0 \left(2 i \mu -\sqrt{\vec{q}^2+(q_0-2 i \mu
   )^2}\right)}{\vec{q}^2+q_0 (q_0-4 i \mu )}\right)\right]\right\}
\end{align}

\subsection{$q_0=0$}

In the static,  $q_0\to 0$ limit of \eqref{integratedchi} and \eqref{integratedxi} we can get the static forms of $\chi(q)$ and $\xi(q)$ (where we add back
the zero $\mu$ expressions for these, which can be deduced from the computation in the previous appendix).
For $\chi(q)$ we get
\begin{equation}
\chi(q)\big |_{q_0=0}=\left\{\begin{array}{lr}
        \dfrac{e^2N\mu}{2 \pi q^2}+ \dfrac{e^2 N}{16 q}, & \text{for } q\leq 2\mu\\
          \dfrac{e^2N\mu}{2 \pi q^2} +\dfrac{e^2 N}{8 q} -\dfrac{e^2N\mu}{4 \pi q^3}\sqrt{q^2-4\mu^2}-\dfrac{e^2N\mu}{8 \pi q}{\rm tan^{-1}}\dfrac{2\mu}{\sqrt{q^2-4\mu^2}}, & \text{for } q\geq 2\mu
        \end{array}\right.
\end{equation}
 
For $\xi(q)$ we get
\begin{equation}
\xi(q)\big |_{q_0=0}=\left\{\begin{array}{lr}
         \dfrac{3e^2 N}{16 q}, & \text{for } q\leq 2\mu\\
        \dfrac{e^2 N}{8 q}   -\dfrac{e^2N\mu}{4 \pi q^3}\sqrt{q^2-4\mu^2}+\dfrac{e^2N\mu}{8 \pi q}{\rm tan^{-1}}\dfrac{2\mu}{\sqrt{q^2-4\mu^2}}, & \text{for } q\geq 2\mu
        \end{array}\right.
\end{equation}
  
\subsection{Small $q$ limit}

At the first few orders in the $q\to 0$ limit $\chi(q)$ and $\xi(q)$ can be obtained by first expanding for small $q$ \eqref{chi} and \eqref{xi} and then integrating in $p$. By doing this for $\chi(q)$ we get
\begin{equation}
\chi(q)=\frac{e^2 N }{8 \pi  q_0^2}\left[q_0\tan ^{-1}\left(\frac{2 \mu }{q_0}\right)-2 \mu
   \right]+\frac{e^2 N q^2 }{16 \pi }\left[\frac{3 \mu
   }{q_0^4}-\frac{\mu }{\left(4 \mu^2+q_0^2\right)^2}-\frac{\tan ^{-1}\left(\frac{2 \mu
   }{q_0}\right)}{q_0^3}\right]+O\left(q^4\right)
\end{equation}
and for $\xi(q)$
\begin{align}
\xi(q)&=\frac{3 e^2 N }{8 \pi  q^2}\left[q_0\tan ^{-1}\left(\frac{2 \mu }{q_0}\right)-2 \mu
   \right]-\frac{e^2 N}{16 \left(\pi 
   \left(q_0^3+4 \mu ^2 q_0\right)^2\right)} \left[q_0\left(4 \mu ^2+q_0^2\right)^2
   \tan ^{-1}\left(\frac{2 \mu }{q_0}\right)
       \right.\nonumber\\&\left.
   -2 \mu  \left(56 \mu
   ^4+q_0^4+28 \mu ^2 q_0^2\right)\right]+O\left(q^2\right)
   \end{align}

\acknowledgments
\noindent
This work was supported in part by NSERC and the INFN. G.G. acknowledge support from the project "Black holes, neutron stars and gravitational waves"  financed by Fondo Ricerca di Base 2018 of the University of Perugia. G.G. acknowledges the hospitality of
the University of British Columbia where this work was begun.  This work was  performed in part at Aspen Center for Physics, 
which is supported by National Science Foundation grant PHY-1607611.


  \end{document}